\documentclass[journal]{IEEEtran}
\usepackage{amsmath}
\usepackage{amsfonts, amssymb}
\usepackage{algorithm}
\usepackage{algorithmic}
\usepackage{graphicx}
\usepackage{epstopdf}
\usepackage{subfigure}
\usepackage{booktabs}
\usepackage{stfloats}
\usepackage{multirow}
\usepackage[noadjust]{cite}
\usepackage{tabularx}
\usepackage{amsmath} 
\usepackage{amssymb}  
\usepackage{hyperref}

\begin{document}
\title{Maximum Total Correntropy Diffusion Adaptation over Networks with Noisy Links}
%
\author{
Yicong He, Fei Wang, \IEEEmembership{Member,~IEEE}, Shiyuan Wang, \IEEEmembership{Member,~IEEE}, Pengju Ren, \IEEEmembership{Member,~IEEE}, Badong Chen, \IEEEmembership{Senior~Member,~IEEE}\thanks{This work was supported by the National Key Research and Development Program of China under grant 2017YFB1002501, National Natural Science Foundation of China under grant 61603291, the program of introducing talents of discipline to university B13043, and also supported in part by the National Natural Science Foundation-Shenzhen Joint Research Program (U1613219).} \thanks{Yicong He, Fei Wang, Pengju Ren and Badong Chen are with the Institute of Artificial Intelligence and Robotics, Xi'an Jiaotong University, China, e-mails: heyicong@stu.xjtu.edu.cn, wfx@xjtu.edu.cn, pengjuren@gmail.com, chenbd@xjtu.edu.cn.} \thanks{ Shiyuan Wang is with the School of Electronic and Information Engineering, Southwest University, China, e-mail: wsy@swu.edu.cn}}
\IEEEpeerreviewmaketitle
\maketitle
%
%

%
%
\begin{abstract}
Distributed estimation over networks draws much attraction in recent years. In many situations, due to imperfect information communication among nodes, the performance of traditional diffusion adaptive algorithms such as the diffusion LMS (DLMS) may degrade. To deal with this problem, several modified DLMS algorithms have been proposed. However, these DLMS based algorithms still suffer from biased estimation and are not robust to impulsive link noise. In this paper, we focus on improving the performance of diffusion adaptation with noisy links from two aspects: accuracy and robustness. A new algorithm called diffusion maximum total correntropy (DMTC) is proposed. The new algorithm is theoretically unbiased in Gaussian noise, and can efficiently handle the link noises in the presence of large outliers. The adaptive combination rule is applied to further improve the performance. The stability analysis of the proposed algorithm is given. Simulation results show that the DMTC algorithm can achieve good performance in both Gaussian and non-Gaussian noise environments.
\end{abstract}


\section{Introduction}
\label{sec:intro}
The goal of the distributed estimation is to estimate the system parameters across the network based on observed data of all nodes. 
Diffusion adaptation has been intensively studied during the last decades \cite{chen2012diffusion,sayed2013diffusion,lopes2008diffusion,cattivelli2010diffusion}. Most algorithms assume that the information exchange among neighbour nodes is noiseless, which, however may not satisfy in real applications. So far, there is a lot of researches on performance analysis of diffusion adaptation with noisy links \cite{Zhao2012Diffusion,Khalili2012Steady,Abdolee2011Diffusion}. However, most studies focus only on the impact of noise on the diffusion process, but give no way to improve the performance of the algorithms.  

Recently, some improved diffusion adaptation algorithms have been proposed to alleviate the impact of noisy communications \cite{Zhao2012Diffusion,Zhao2012Clustering,Nassif2016Diffusion}. By modifying the diffusion strategies of typical diffusion least mean squares (DLMS) algorithm, these algorithms can work well under noisy links. These algorithms utilize mean squared error (MSE) as the cost function, however, many lead to biased estimation. In particular, MSE can obtain unbiased estimation for standard regression assumption where input data is noiseless. While for noisy communications, all the data are perturbed by noise during information exchange process. This recalls the error in variable (EIV) model which takes both input and output data into consideration. So far several diffusion algorithms under EIV model have been proposed such as bias-compensated DLMS (BC-DLMS) algorithm \cite{abdolee2016diffusion} and diffusion gradient descent based total least squares (D-GDTLS) algorithm \cite{arablouei2013diffusion,Huang2015Distributed}. 
Both methods can achieve better performance compared with DLMS for noisy input and output data. 

The robustness is also an important issue in real applications. BC-DLMS and D-GDTLS are all based on quadratic form costs, which can perform well in Gaussian noises. However, due to impulsive perturbation, the noise distribution may contain large outliers, thus the performance of quadratic cost based algorithms may degrade. Recently, an information theoretic learning (ITL) based criterion called maximum total correntropy (MTC) has been proposed \cite{Wang2017Maximum}. Correntropy is a nonlinear method to measure the similarity, which involves all the even moments of the difference and is insensitive to outliers \cite{correntropy}. As an extension of maximum correntropy criterion (MCC) \cite{correntropy,chen2014steady,chen2012maximum,wu2015robust,chen2015convergence,Ma2016Diffusion} and total least squares (TLS) \cite{tlsoverview}, MTC aims to deal with the robust EIV regression problem. MTC based adaptive filtering has shown accurate and robust performance when both input and output noise contain large outliers \cite{correntropy}.

In this paper we develop a new algorithm called diffusion maximum total correntropy (DMTC) algorithm for noisy links. Using EIV model, DMTC can theoretically produce unbiased estimation in Gaussian noise. Taking advantage of correntropy, DMTC can efficiently handle the large outliers. The theoretical analysis in the mean and mean-square error performance is also given. Moreover, to reduce the negative influence of noise in combination step, the adaptive combination rule is utilized. Simulation results confirm the desired performance of the proposed algorithm, showing that our method can greatly improve the performance of diffusion adaptation with noisy links in both accuracy and robustness.


\section{Proposed algorithm}

\subsection{System model and problem statement}
\label{sec:format}
Consider a connected network with $N$ nodes. For each node $k$, suppose we are given a sequence $\{\boldsymbol{x}_k(i),d_k(i)\}^M_{i=1}$ with the following linear relation
\begin{equation}
d_k(i) = {\boldsymbol{h}^T}\boldsymbol{x}_k(i)
\end{equation}
where ${\boldsymbol{x}}_k(i)\in R^{L\times1}$ is the input vector of node $k$ at time $i$, $d_k(i)\in R$ is the corresponding output and $\boldsymbol{h}\in R^{L\times1}$ is the target weight vector. Since the output data may be perturbed by measurement noise, the observed data can be described as
\begin{equation}
\begin{aligned}
y_k(i) = d_k(i) + v_k(i)
\end{aligned}
\end{equation}
where $v_k(i)$ is the output noise of node $k$ with variance $\sigma_{k}^2$. The goal of distributed estimation is to collaboratively estimate $\boldsymbol{h}$ based on data sequences $\{{\boldsymbol{x}}_k(i),y_k(i)\}^M_{i=1}, k=1,...,N$.
\par Diffusion adaptation strategies have been widely used in distributed estimation. In particular, DLMS is the most used algorithm. The weight update of DLMS is
\begin{equation}
\label{DLMS}
\left\{ {\begin{array}{*{20}{c}}
{{\boldsymbol{\phi} _k}\left( i+1 \right) = {{\boldsymbol{w}} _k}\left( {i} \right)+{\mu _k}\sum\limits_{l \in {{\cal N}_k}} {{\alpha _{lk}}{{\hat {\boldsymbol{g}}}_{l}}\left( {{{\boldsymbol{w}} _k}\left( {i} \right)} \right)} }\\
{{{\boldsymbol{w}} _k}\left( {i + 1} \right) = \sum\limits_{l \in {{\cal N}_k}} {{\beta _{lk}}{\boldsymbol{\phi}_l}\left( {i + 1} \right)} }
\end{array}} \right.
\end{equation}
where
${{\hat {\boldsymbol{g}}}_{l}}\left( {{{\boldsymbol{w}} _k}\left( {i} \right)} \right)={{( y_l(i) -  {{{\boldsymbol{w}} _k^T}\!\left( {i} \right)}{{\boldsymbol{x}}_l(i)}){{\boldsymbol{x}}_l(i)}}}$,
${\boldsymbol{\phi} _k(i)}$ is the intermediate vectors of node $k$ at time $i$, ${{\cal N}_k}$ is the neighbourhood of node $k$, $\mu_k$ is the corresponding step-size, and $\{\alpha_{lk}\}, \{\beta_{lk}\}$ are the $\{l,k\}$-th entries of matrices $\boldsymbol{A}$ and $\boldsymbol{C}$ respectively. In particular, $\boldsymbol{A}$ and $\boldsymbol{C}$ should satisfy $\boldsymbol{A}^T\boldsymbol{1}=\boldsymbol{C}\boldsymbol{1}=\boldsymbol{1}$, and $\boldsymbol{A}(l,k)=\boldsymbol{C}(l,k)=0$ if $l\notin {\cal N}_k$.
\par Eq.(\ref{DLMS}) contains two information communication procedures. For node $k$ at each iteration $i$, the data $\{{{\boldsymbol{x}}_k(i)},y_k(i)\}$ are shared to neighbour node $l\in{\cal N}_k$ for gradient estimation, and then receive the estimated vector ${\boldsymbol{\phi}_l}\left( {i} \right)$ from neighbour node $l$ for combination. When taking link noise into consideration,
the observed data are represented as
\begin{equation}
\begin{aligned}
y_{lk}(i) = y_{l}(i) + v_{lk,y}(i)\\
{\boldsymbol{x}}_{lk}(i) = {\boldsymbol{x}}_{l}(i) + {\boldsymbol{v}}_{lk,{\boldsymbol{x}}}(i)\\
{\boldsymbol{\phi}}_{lk}(i) = {\boldsymbol{\phi}}_{l}(i) + {\boldsymbol{v}}_{lk,{\boldsymbol{\phi}}}(i)
\end{aligned}
\end{equation}
where $y_{lk}(i),{\boldsymbol{x}}_{lk}(i),{\boldsymbol{\phi}}_{lk}(i)$ are the noisy data received by node $k$ from its neighbor $l$, $v_{lk,y}(i),{\boldsymbol{v}}_{lk,{\boldsymbol{x}}}(i),{\boldsymbol{v}}_{lk,{\boldsymbol{\phi}}}(i)$ are the corresponding measurement noise with variance $\sigma_{lk,y}^2$ and covariance matrix $\sigma_{lk,\boldsymbol{x}}^2 \boldsymbol{I}_{L\times L}$, $\sigma_{lk,\boldsymbol{\phi}}^2 \boldsymbol{I}_{L\times L}$ respectively. Note that $\sigma_{kk,y}^2=\sigma_{kk,\boldsymbol{x}}^2=\sigma_{kk,\boldsymbol{\phi}}^2=0$ since self weight update will not contain noise. In the following part, for simplicity, the subscript $kk$ and $k$ denote the same notation.
\subsection{DMTC algorithm}
It is known that DLMS utilizes the following approximation of global cost function at node $k$ \cite{chen2012diffusion,cattivelli2010diffusion}
\begin{equation}
\label{Jglobal}
J_k^{glob}({\boldsymbol{w}})=\sum_{l\in{\cal N}_k}{\alpha _{lk}}J_{l}({\boldsymbol{w}})+\sum_{l\in{\cal N}_k\backslash\{k\}}b_{lk}\|{\boldsymbol{w}}-{\boldsymbol{h}}\|^2
\end{equation}
where ${b_{lk}}$ can be obtained from ${\beta_{lk}}$, the local cost function $J_l({\boldsymbol{w}})$ at node $l$ is the MSE based cost function
\begin{equation}
\label{JMSE}
J_l({\boldsymbol{w}})={E\left[ {{{( y_l(i) - {{\boldsymbol{w}}^T}{{\boldsymbol{x}}_l(i)})}^2}} \right]}
\end{equation}
When information exchange contains noise, using the MSE as the cost function may lead to biased estimation. In particular, during data sharing from node $l$ to $k$, the input data $\boldsymbol{x}_{k}(i)$ and output data $d_{k}(i)$ are perturbed by noise vector ${\boldsymbol{v}}_{lk,{\boldsymbol{x}}}(i)$ and noise $v_l(i)+v_{lk,y}(i)$, respectively. Therefore, it is more reasonable to treat the regression model for ${l\in{\cal N}_k\backslash\{k\}}$ as the EIV model where all the variables are assumed to be perturbed by noise. For EIV regression in distributed network, The D-GDTLS algorithm has been proposed \cite{arablouei2013diffusion}. Directly applying D-GDTLS to the proposed problem leads to the following cost function
\begin{equation}
\label{DGDTLScost}
J_{lk}(\boldsymbol{w}) = E\left[\frac{{ {{{(y_{lk}(i) - {{\boldsymbol{w}}^T}{{\boldsymbol{x}}_{lk}(i)})}^2}} }}{{{{\left\| {\boldsymbol{w}} \right\|}^2} + \gamma_{lk} }}\right]~,~~{l\in{\cal N}_k\backslash\{k\}}
\end{equation}
where $\gamma_{lk}=(\sigma_{l}^2+\sigma_{lk,y}^2)/\sigma_{lk,\boldsymbol{x}}^2$. Here we use the notation $J_{lk}(\boldsymbol{w})$ instead of $J_{l}(\boldsymbol{w})$ in Eq.(\ref{Jglobal}) since cost functions for each link may no longer the same.

D-GDTLS uses quadratic form cost, thus may not perform well when link noise contains large outliers. In adaptive filtering, to improve the robustness of EIV model regression, an ITL based criterion called maximum total correntropy (MTC) is developed \cite{Wang2017Maximum}. By replacing the quadratic norm in Eq.(\ref{DGDTLScost}) with correntropy measure, one can obtain the MTC based utility functions
\begin{equation}
\label{JMTC}
J_{lk}(\boldsymbol{w})\!=\!E\!\left[ {\exp \left( { - \frac{{{{\left(  {y}_{lk}(i) - {{\boldsymbol{w}}^T}{{{\boldsymbol{x}}}_{lk}(i)} \right)}^2}}}{{2\zeta _{lk}^2\left( {{{\left\| {\boldsymbol{w}} \right\|}^2} + {\gamma _{lk}}} \right)}}} \right)} \right],~ {l\in{\cal N}_k\backslash\{k\}}
\end{equation}
where $\zeta _{lk}$ is the kernel parameter and $\gamma_{lk}$ is calculated without outliers. 
For self weight update (i.e. $l=k$), the MSE in Eq.(\ref{JMSE}) is still used as the cost function. Therefore, based on Eq.(\ref{JMSE}) and Eq.(\ref{JMTC}), we derive the diffusion maximum total correntropy (DMTC) algorithm for noisy communications as
\begin{equation}
\label{DMTC}
\left\{ {\begin{array}{*{20}{c}}
{{\boldsymbol{\phi} _k}\left( i+1 \right) = {{\boldsymbol{w}} _k}\left( {i} \right)+{\mu _k}\sum\limits_{l \in {{\cal N}_k}} {{\alpha _{lk}}{{\hat {\boldsymbol{g}}}_{lk}}\left( {{{\boldsymbol{w}} _k}\left( {i} \right)} \right)} }\\
{{{\boldsymbol{w}} _k}\left( {i + 1} \right) = \sum\limits_{l \in {{\cal N}_k}} {{\beta_{lk}}{\boldsymbol{\phi}_{lk}}\left( {i + 1} \right)} }
\end{array}} \right.
\end{equation}
where 
\begin{equation*}
\begin{aligned}
\label{MTC}
&{\hat {\boldsymbol{g}}}_{lk}(\boldsymbol{w}_k(i))\\ 
&=\begin{cases}
e_{lk}(i){\boldsymbol{ x}}_{lk}(i)~,~~ l=k\\
\!G_{lk}(i)\!\frac{{ {{({\left\| {\boldsymbol{w}_k\!(i)} \right\|}^2\!\!+\!\gamma_{lk})} } e_{lk}(i){\boldsymbol{x}}_{lk}(i) \!+\! {e_{lk}^2(i)}{\boldsymbol{w}_k(i)}}}{{ {\zeta_{lk}^2({\left\| {\boldsymbol{w}_k\!(i)} \right\|^2+\gamma_{lk}})^2} }},{l\!\in\!{\cal N}_k\backslash\{k\}}
\end{cases}
\end{aligned}
\end{equation*}
\begin{equation*}
\begin{aligned}
&e_{lk}(i)={y_{lk}(i) - {{\boldsymbol{w}_k^T(i)}}{{\boldsymbol{x}}_{lk}(i)}}\\
&G_{lk}(i)=\exp\! \left( \!{ - \frac{{{e_{lk}^2}(i)}}{{2\zeta_{lk}^2 {({{\left\| {\boldsymbol{w}_k(i)} \right\|}^2+\gamma_{lk})}} }}} \!\right)\!
\end{aligned}
\end{equation*}
Note that the coefficient $1/{{\zeta _{lk}^2}}$ in the derivation of Eq.(\ref{JMTC}) is absorbed into $\mu_k$.

\subsection{Adaptive combination rule}
\label{sec:pagestyle}
To further improve the performance and robustness, we follow the adaptive combination rule \cite{Zhao2012Diffusion,Nassif2016Diffusion} and propose the adaptive combination DMTC (AC-DMTC) algorithm. In particular, the coefficient $\beta_{lk}$ is replaced by
\begin{equation}
\label{acrule}
\beta_{lk}(i+1)=
\begin{cases}
\frac{\delta_{lk}^{-2}(i+1)}{\sum_{j\in{\cal N}_k}{\delta_{jk}^{-2}(i+1)}},l\in{\cal N}_k\\
0,l\notin{\cal N}_k
\end{cases}
\end{equation}
where 
\begin{equation}
\label{smooth}
\delta_{lk}^{2}(i+1)=(1-\chi)\delta_{lk}^{2}(i)+\chi\|\boldsymbol{\phi}_{lk}(i+1)-\boldsymbol{\hat w}_k(i+1)\|^2
\end{equation}
is a smoothed estimation of $\delta_{lk}^{2}(i+1)$ with a forgetting factor $\chi$, and $\boldsymbol{\hat w}_k(i+1)$ is a local one-step approximation defined as \cite{Nassif2016Diffusion}
\begin{equation}
\boldsymbol{\hat w}_k(i+1)=\boldsymbol{w}_k(i)+\mu_{k}\frac{{{\hat {\boldsymbol{g}}}_{k}}(\boldsymbol{w}_k(i))}{\|{{{\hat {\boldsymbol{g}}}_{k}}(\boldsymbol{w}_k(i))}\|^2+\epsilon}
\end{equation}
where $\epsilon$ is a sufficient small positive value.
\par One should notice that this adaptive combination rule is essentially robust to outliers. In particular, using smoothed estimation in Eq.(\ref{smooth}) reduces the negative impact of accidental large outliers. Moreover, the assignment in Eq.(\ref{acrule}) also guarantee the robustness since relative large $\delta_{lk}^{2}(i+1)$ will result small $\beta_{lk}(i+1)$.

\section{Local Convergence analysis}
\label{sec:typestyle}
In this section, we carry out the local convergence analysis of the proposed DMTC algorithm. The analysis is based on the following assumption:
\par $Assumption~1$: For arbitrary node $k$, the elements of $v_k(i),v_{lk,y}(i),{\boldsymbol{v}}_{lk,{\boldsymbol{x}}}(i),{\boldsymbol{v}}_{lk,{\boldsymbol{\phi}}}(i)$ are zero-mean i.i.d. Gaussian random processes, and are independent of ${\boldsymbol{x}}_k(i)$ and $d_k(i)$.
\par The above assumption is commonly used in analysis of diffusion algorithms \cite{chen2012diffusion,sayed2013diffusion,lopes2008diffusion,cattivelli2010diffusion,Zhao2012Diffusion}. 
Since calculation in non-Gaussian noise is very hard, the strictly analysis of the proposed algorithm is based on Gaussian noise assumption. The performance on non-Gaussian link noise is briefly discussed and will be verified by simulations. Moreover, similar to \cite{Zhao2012Diffusion,Nassif2016Diffusion}, during the analysis we regard $\boldsymbol{C}$ as a fixed matrix.
\subsection{Mean stability}
First we analyze the mean stability of the proposed algorithm. 
Subtracting both sides of Eq.(\ref{DMTC}) from $\boldsymbol{h}$ we can obtain
\begin{equation}
\label{DMTCh}
\left\{ {\begin{array}{*{20}{c}}
{{\tilde{\boldsymbol{\phi}} _k}\left( i+1 \right) = {\tilde{\boldsymbol{w}} _k}\left( {i} \right)-{\mu _k}\sum\limits_{l \in {{\cal N}_k}} {{\alpha _{lk}}{{\hat {\boldsymbol{g}}}_{lk}}\left( {{\boldsymbol{w} _k}\left( {i} \right)} \right)} }\\
{{\tilde{\boldsymbol{w}}_k}\left( i+1 \right) = \sum\limits_{l \in {{\cal N}_k}} {{\beta _{lk}}\left({\tilde{\boldsymbol{\phi}} _l}\left( i+1 \right)-{\boldsymbol{v}}_{lk,{\boldsymbol{\phi}}}(i+1)\right)} }
\end{array}} \right.
\end{equation}
where ${\tilde{\boldsymbol{w}} _k}\left( {i} \right) = \boldsymbol{h} - {{\boldsymbol{w}} _k}(i)$, $\tilde {\boldsymbol{w}}_k(i) = \boldsymbol{h} - \boldsymbol{w}_k(i)$, ${\tilde{\boldsymbol{\phi}} _k}\left( i \right) = \boldsymbol{h} - \boldsymbol{\phi}_k(i)$ are error vectors. For further analysis, we define instantaneous gradient error vector ${\boldsymbol{s}_{lk}}({{\boldsymbol{w} _k}\left( {i} \right)})$ as
\begin{equation}
\label{EMTC}
{\boldsymbol{s}_{lk}}({{\boldsymbol{w} _k}\left( {i} \right)})={{\hat {\boldsymbol{g}}}_{lk}}\left( {{{\boldsymbol{w}} _k}\left( {i} \right)} \right) - {\boldsymbol{g}_{lk}}\left( {{\boldsymbol{w} _k}\left( {i} \right)} \right)
\end{equation}
where ${\boldsymbol{g}_{lk}}\left( {{\boldsymbol{w} _k}\left( {i} \right)} \right)=E[{{\hat {\boldsymbol{g}}}_{lk}}\left( {{{\boldsymbol{w}} _k}\left( {i} \right)} \right)]$. Since ${J_{lk}} (\boldsymbol{w})$ is twice continuously differentiable in a neighborhood of a line segment between points $\boldsymbol{w}_k (i)$ and $\boldsymbol{h}$, we can use Theorem 1.2.1 in \cite{Kelley2010Iterative} and approximate ${\boldsymbol{g}}_{lk}\left( {{\boldsymbol{w} _k}\left( {i} \right)} \right)$ as \cite{chen2012diffusion} 
\begin{equation}
\begin{aligned}
\label{EMTC4}
{\boldsymbol{g}_{lk}} (\boldsymbol{w}_k (i)) &= {\boldsymbol{g}_{lk}} (\boldsymbol{h}) - \left[ {\int_0^1 {{\boldsymbol{H}_{lk}}\left ( {\boldsymbol{h} - t {\tilde{\boldsymbol{w}}_k} (i)} \right)dt} } \right] {\tilde{\boldsymbol{w}}_k} (i)\\
& \approx {\boldsymbol{g}_{lk}} (\boldsymbol{h})- \left[ {\int_0^1 {{\boldsymbol{H}_{lk}}\left ( \boldsymbol{h} \right)dt} } \right] {\tilde{\boldsymbol{w}}_k} (i)\\
& \approx  {\boldsymbol{g}_{lk}} (\boldsymbol{h}) - {\boldsymbol{H}_{lk}}\left ( \boldsymbol{h} \right) {\tilde {\boldsymbol{w}}_k} (i)
\end{aligned}
\end{equation}
where ${\boldsymbol{H}_{lk}}\left ( \boldsymbol{h} \right)$ is the Hessian matrix of $J_{lk}(\boldsymbol{w})$ at $\boldsymbol{h}$.
\par For Gaussian noise distribution, using integral method to compute the expectation, one can obtain \cite{Wang2017Maximum}
\begin{equation}
{\boldsymbol{g}_{lk}} (\boldsymbol{h})=\boldsymbol{0}~,~ l\in{\cal N}_k
\end{equation}
\begin{equation}
\label{gH}
{{\boldsymbol{H}}_{lk}({\boldsymbol{h}})}=
\begin{cases}
- {\boldsymbol{R}_l},l=k\\
- \frac{1}{{{{\left\| {\boldsymbol{h}} \right\|}^2+\gamma_{lk}}}}{{\left(\frac{{\zeta _{lk}^2}}{{\sigma_{lk,\boldsymbol{x}}^2 + \zeta _{lk}^2}}\right)}^{\frac{3}{2}}}{\boldsymbol{R}_l},{l\!\in\!{\cal N}_k\backslash\{k\}}
\end{cases}
\end{equation}
where $\boldsymbol{R}_l = E[{\boldsymbol{x}}_l(i){\boldsymbol{x}}_l^T(i)]$ is the covariance matrix of node $l$. Then, by defining
\begin{gather*}
\boldsymbol{\cal A}=\boldsymbol{A} \otimes \boldsymbol{I}_L,\tilde {\boldsymbol{w}}(i)={\rm{col}}\{\tilde {\boldsymbol{w}}_p(i)\}_{p=1}^N,\boldsymbol{\cal M}={\rm diag}\{\mu_p\boldsymbol{I}_L\}_{p=1}^N\\
\boldsymbol{\cal H} =\sum\nolimits_{l = 1}^N {\rm diag}\{\alpha_{lp}{\boldsymbol{H}_{lp}}\left( \boldsymbol{h} \right)\}_{p=1}^N\\
\boldsymbol{s}(i)=\sum\nolimits_{l = 1}^N {\rm col}\{\alpha_{lp}{\boldsymbol{s}_{lp}(\boldsymbol{w} _p(i))}\}_{p=1}^N\\
\boldsymbol{z}(i)=\sum\nolimits_{l = 1}^N {\rm col}\{{\beta _{l1}}{\boldsymbol{v}}_{l1,{\boldsymbol{\phi}}}(i),...,{\beta _{lN}}{\boldsymbol{v}}_{lN,{\boldsymbol{\phi}}}(i)\}
\end{gather*}
and substituting Eq.(\ref{EMTC})-(\ref{gH}) into Eq.(\ref{DMTCh}) yields
\begin{equation}
\label{EMTC3}
\tilde {\boldsymbol{w}}(i + 1) \!= \!\boldsymbol{\cal A}^T[\boldsymbol{I}+\boldsymbol{\cal M}\boldsymbol{\cal H}]\tilde {\boldsymbol{w}}(i) \!- \!\boldsymbol{\cal A}^T\boldsymbol{\cal M}\boldsymbol{s}(i)\!-\!\boldsymbol{z}(i+1)
\end{equation}
Under Assumption 1 and taking expectation of both sides, we get
\begin{equation}
\label{EMTC5}
E[\tilde {\boldsymbol{w}}(i + 1)] = \boldsymbol{\cal A}^T[\boldsymbol{I}+\boldsymbol{\cal M}\boldsymbol{\cal H}] E[\tilde {\boldsymbol{w}}(i)]
\end{equation}
Therefore, to ensure the stability, $\boldsymbol{\cal A}^T[\boldsymbol{I}+\boldsymbol{\cal M}\boldsymbol{\cal H}]$ should be stable, i.e. $\rho(\boldsymbol{\cal A}^T[\boldsymbol{I}+\boldsymbol{\cal M}\boldsymbol{\cal H}])<1$ where $\rho(.)$ is the spectrum operator. Since $\boldsymbol{A}^T\boldsymbol{1}=\boldsymbol{1}$, we obtain $\rho(\boldsymbol{\cal A})=1$. Thus the magnitudes of all the eigenvalues of matrix $\boldsymbol{I}+\boldsymbol{\cal M}\boldsymbol{\cal H}$ must be less than unity to ensure the local convergence. After some algebra, one can conclude that the step size $\mu_k$ should satisfy the following condition to ensure the convergence
\begin{equation}
\label{stepsize2}
0 < \mu_k  < \min \left\{ \frac{{2}}{\rho\left(\sum_{l=1}^{N}\alpha_{lk}\boldsymbol{H}_{lk}(\boldsymbol{h}) \right)} \right\}
\end{equation}
\par When Eq.(\ref{stepsize2}) is satisfied, $E[\tilde {\boldsymbol{w}}(i)]$ will tend to zero as $i\rightarrow\infty$, that is, under Gaussian noise, the estimation of the proposed DMTC algorithm is asymptotically unbiased. 

\textbf{Remark 1}: When Gaussian noise is contaminated with occasional large outliers, due to utilizing Gaussian function, large outliers will have little influence on the local maximum property of local cost function $J_{lk}({\boldsymbol{w}})$. Therefore, DMTC can efficiently alleviate the bad influence of large outliers. Moreover, since MTC is only the local maximum, to guarantee a global optimal solution, one can use D-GDTLS or DMTC with sufficient large $\zeta_{lk}$ to train the adaptive filter first to make sure the solution is near the global optimal solution \cite{Wang2017Maximum}.

\subsection{Mean square stability and steady state performance}
At the vicinity of local maximum, the gradient error ${\boldsymbol{s}_{lk}}({{\boldsymbol{w} _k}\left( {i} \right)})$ in Eq.({\ref{EMTC}}) can be approximated as
\begin{equation}
{\boldsymbol{s}_{lk}} (\boldsymbol{w} (i))\approx {\boldsymbol{s}_{lk}} (\boldsymbol{h}) ={\hat {\boldsymbol{g}}_{lk}} (\boldsymbol{h})
\end{equation}
Thus squaring both sides of Eq.({\ref{EMTC3}}) gives the following mean energy conservation \cite{chen2012diffusion}
\begin{equation}
E\left[ {\left\| {\tilde {\boldsymbol{w}} (i + 1)} \right\|_{\boldsymbol{\Sigma}}^2} \right] \approx E\left[ {\left\| {\tilde {\boldsymbol{w}} (i)} \right\|_{\boldsymbol{\Lambda}} ^2} \right] + tr\{{\boldsymbol{\Sigma}}\left({{\boldsymbol{\cal R}}}+ {{\boldsymbol{\cal V}}}\right)\}
\end{equation}
where ${\boldsymbol{\Sigma}}$ is an arbitrary positive symmetric matrix, $\left\| \boldsymbol{m} \right\|_{\boldsymbol{B}}^2$ denotes ${\boldsymbol{m}^T}\boldsymbol{B}\boldsymbol{m}$ for arbitrary matrix $\boldsymbol{B}$ and
$$\boldsymbol{\Lambda}= [\boldsymbol{I}+\boldsymbol{\cal M}\boldsymbol{\cal H}] \boldsymbol{\cal A}\boldsymbol{\Sigma} \boldsymbol{\cal A}^T[\boldsymbol{I}+\boldsymbol{\cal M}\boldsymbol{\cal H}] ,{\boldsymbol{\cal R}}=\boldsymbol{\cal A}^T\boldsymbol{\cal M}\boldsymbol{\cal C}\boldsymbol{\cal M}^T\boldsymbol{\cal A}$$
$${\boldsymbol{\cal C}}\!=\!\sum_{l = 1}^N\! {\rm diag}\{{\alpha _{lp}^2}{\boldsymbol{Q}_{lp}}\!\left( \boldsymbol{h} \right)\}_{p=1}^N,\boldsymbol{\cal V}\!=\!\sum_{l = 1}^N \!{\rm diag}\{{\beta _{lp}^2}\sigma_{lp,{\boldsymbol{\phi}}}^2\boldsymbol{I}_L\}_{p=1}^N$$
\begin{equation}
\begin{aligned}
\!&{\boldsymbol{Q}_{lk}}\left( \boldsymbol{h} \right) = E\left [ {{{\hat {\boldsymbol{g}}}_{lk}} (\boldsymbol{h})\hat {\boldsymbol{g}}_{lk}^T (\boldsymbol{h})} \right]\\
&= \begin{cases}
 \!{\sigma _{l}^2}{\boldsymbol{R}_l},l=k\\
 \!\frac{{\sigma _{lk,\boldsymbol{x}}^2}}{({{\left\| {\boldsymbol{h}} \right\|}^2+\gamma_{lk}})^2}{{\left (\! {\frac{{\zeta_{lk}^2}}{{2\sigma _{lk,\boldsymbol{x}}^2 + \zeta_{lk}^2}}}\! \right)\!}^{\frac{3}{2}}}\!\!\left[ {{\left({{\left\| {\boldsymbol{h}} \right\|}^2+\gamma_{lk}}\right)}}\right.\\
 ~~~~~~\times\left.{\left ( {{\boldsymbol{R}_l} \!+\! \sigma _{lk,\boldsymbol{x}}^2{\boldsymbol{I}}} \right) \!-\! \sigma _{lk,\boldsymbol{x}}^2{\boldsymbol{h}}{{\boldsymbol{h}}^T}}\! \right],{l\!\in\!{\cal N}_k\backslash\{k\}}
\end{cases}
\end{aligned}
\end{equation}
\par After some algebra, one can obtain the following mean-square relation
\begin{equation}
\label{EMTC7}
E\left[ {\left\| {\tilde {\boldsymbol{w}} (i + 1)} \right\|_{\boldsymbol{\theta}}^2} \right] \approx E\left[ {\left\| {\tilde {\boldsymbol{w}} (i)} \right\|_{\boldsymbol{{\cal F}\theta}} ^2} \right] + vec\{\boldsymbol{\cal R}+\boldsymbol{\cal V}\}^T\boldsymbol{\theta}
\end{equation}
where $\boldsymbol{\theta}= vec\{\boldsymbol{\Sigma}\}$ and
\begin{equation}
\begin{aligned}
\boldsymbol{\cal F} &= [\boldsymbol{I}+\boldsymbol{\cal M}\boldsymbol{\cal H}] \boldsymbol{\cal A}\otimes[\boldsymbol{I}+\boldsymbol{\cal M}\boldsymbol{\cal H}] \boldsymbol{\cal A}
\end{aligned}
\end{equation}
where $vec \{\cdot\}$ is the matrix vectorization operator. ${\left\| {\tilde {\boldsymbol{w}} (i + 1)} \right\|_{\boldsymbol{\theta}}^2}$ and ${\left\| {\tilde {\boldsymbol{w}} (i + 1)} \right\|_{\boldsymbol{\Sigma}}^2}$ denotes the same quantities.
When Eq.({\ref{stepsize2}}) is satisfied, all eigenvalues of $\boldsymbol{\cal F}$ will be also less than $1$, and the mean square stability in Eq.({\ref{EMTC7}}) can be guaranteed. Thus the condition of step-size in Eq.({\ref{stepsize2}}) will guarantee both the mean and mean-square stability.
\par When the iteration number is large enough ($i\rightarrow\infty$), 
we can obtain the steady state MSD as
\begin{equation}
\begin{aligned}
\label{EMTC8}
MSD&\triangleq \lim_{i\rightarrow\infty} \frac{1}{N}\sum\nolimits_{k=1}^{N}E\|\boldsymbol{w}_k(i)\|^2\\
&\approx \frac{1}{N}{ {vec\left\{  \boldsymbol{\cal V}+\boldsymbol{\cal R}\right\}} ^T}{\left ( {\boldsymbol{I} - \boldsymbol{F}} \right)^{ - 1}}vec\left\{ \boldsymbol{I} \right\}
\end{aligned}
\end{equation}

\textbf{Remark 2}: One can observe that the formulation of Eq.(\ref{EMTC8}) is similar to MSD derived for DLMS without data sharing (i.e. $\boldsymbol{A}=\boldsymbol{I}$) \cite{Zhao2012Diffusion}. Thus, similar to analysis in \cite{Zhao2012Diffusion}, one should minimize the upper bound of MSD ${\rm{Tr}\left(  \boldsymbol{\cal V}+\boldsymbol{\cal R}\right)}$ to achieve better performance. This fact theoretically verifies the rationality of using the adaptive combination rule for DMTC.

\begin{figure}[tb]
\centering
\includegraphics[width=0.85\linewidth]{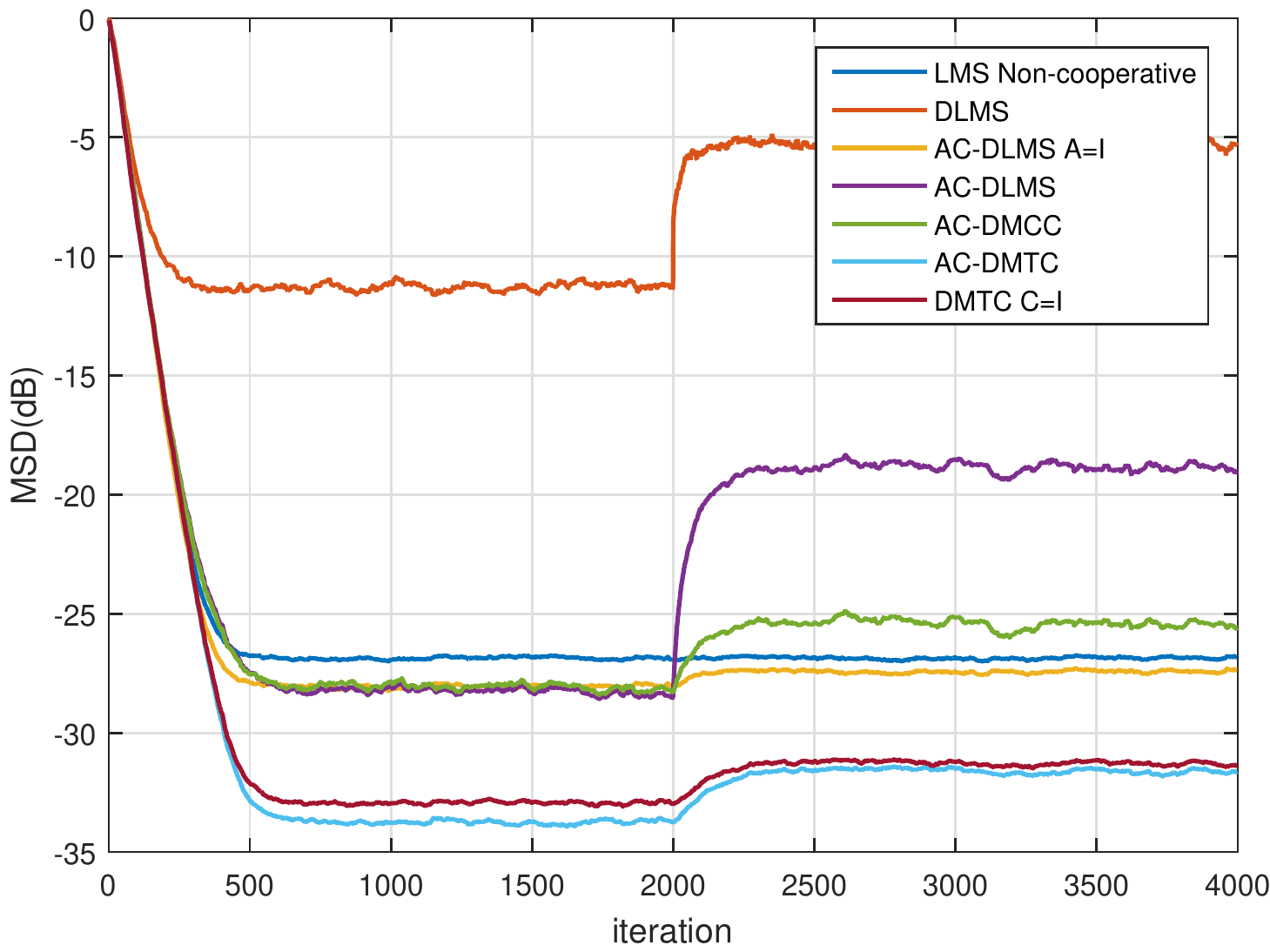}
\caption{Average learning curves in Gaussian and GMM noises}
\label{fig1}
\end{figure}

\section{Simulation Results}
\label{sec:majhead}

In this section, we present simulation results to illustrate the performance of the proposed algorithm. Consider a distributed network with 20 nodes and the desired weight vector is assumed to be
$\boldsymbol{h} = [0.4,0.7,-0.3,0.5]^T$. Each node is linked to 3 nodes in average. Each element of input vectors is zero-mean Gaussian with unit variance.
The Gaussian mixture model (GMM) is used as the distribution of link noise. The probability density function (PDF) of the GMM noise $v$ is $p(v)=(1-c){\cal N}(0,\sigma_A^2)+c{\cal N}(0,\sigma_B^2)$, $\sigma_A^2$ stands for original noise and is associated with calculation of $\gamma_{lk}$. $\sigma_B^2$ is set as large value to represent the distribution of large outliers. The parameter $c$ controls the occurrence probability of large outliers.
%
%

\begin{figure*}[htbp]
\centering
\subfigure
[Gaussian link noise with different $\sigma_A^2$]{
\includegraphics[width=0.28\linewidth]{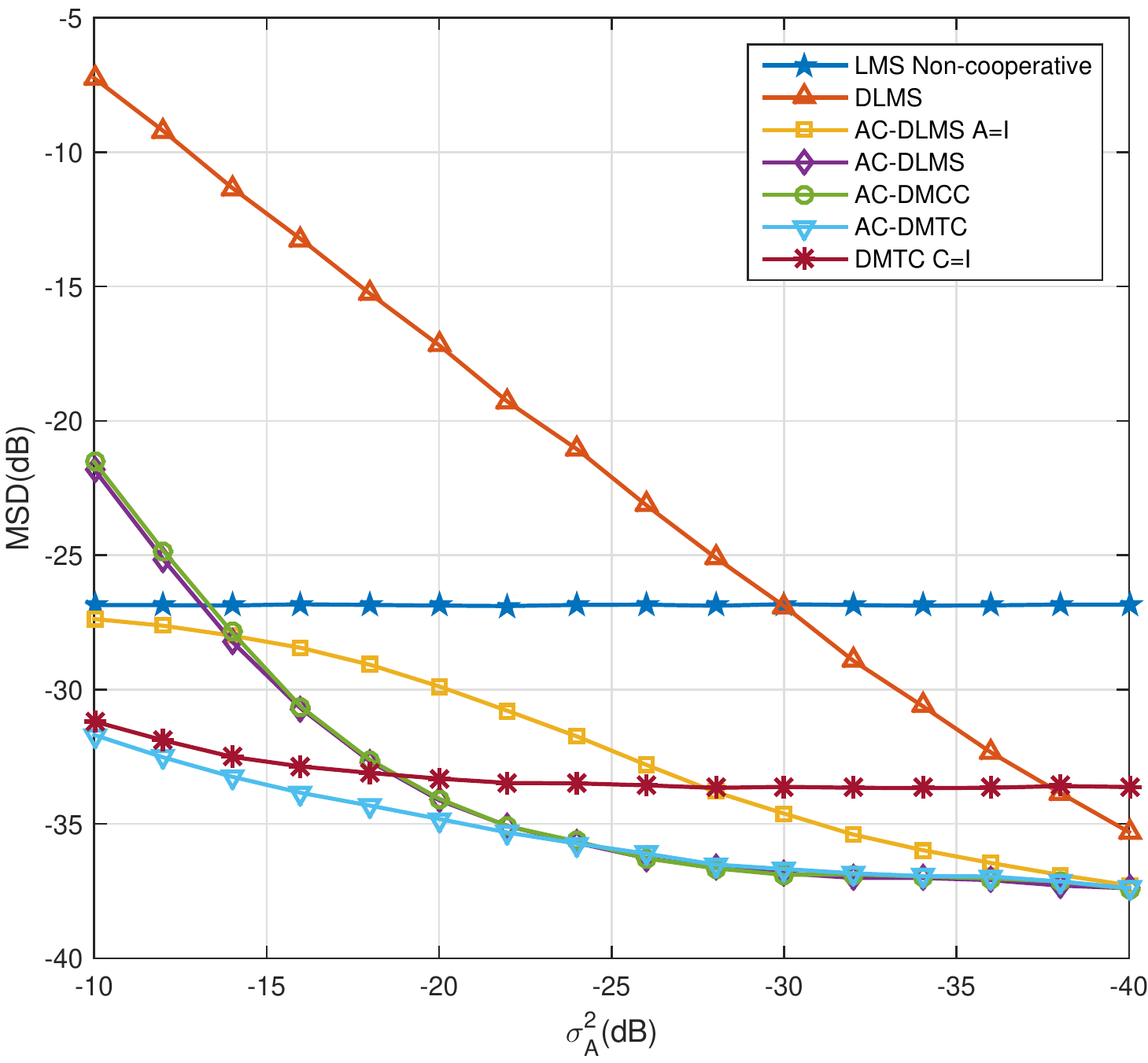}\hspace{0.15in}}
\subfigure[GMM link noise with different $\sigma_A^2$ ($\sigma_B^2=10,c=0.01$)]{
\includegraphics[width=0.28\linewidth]{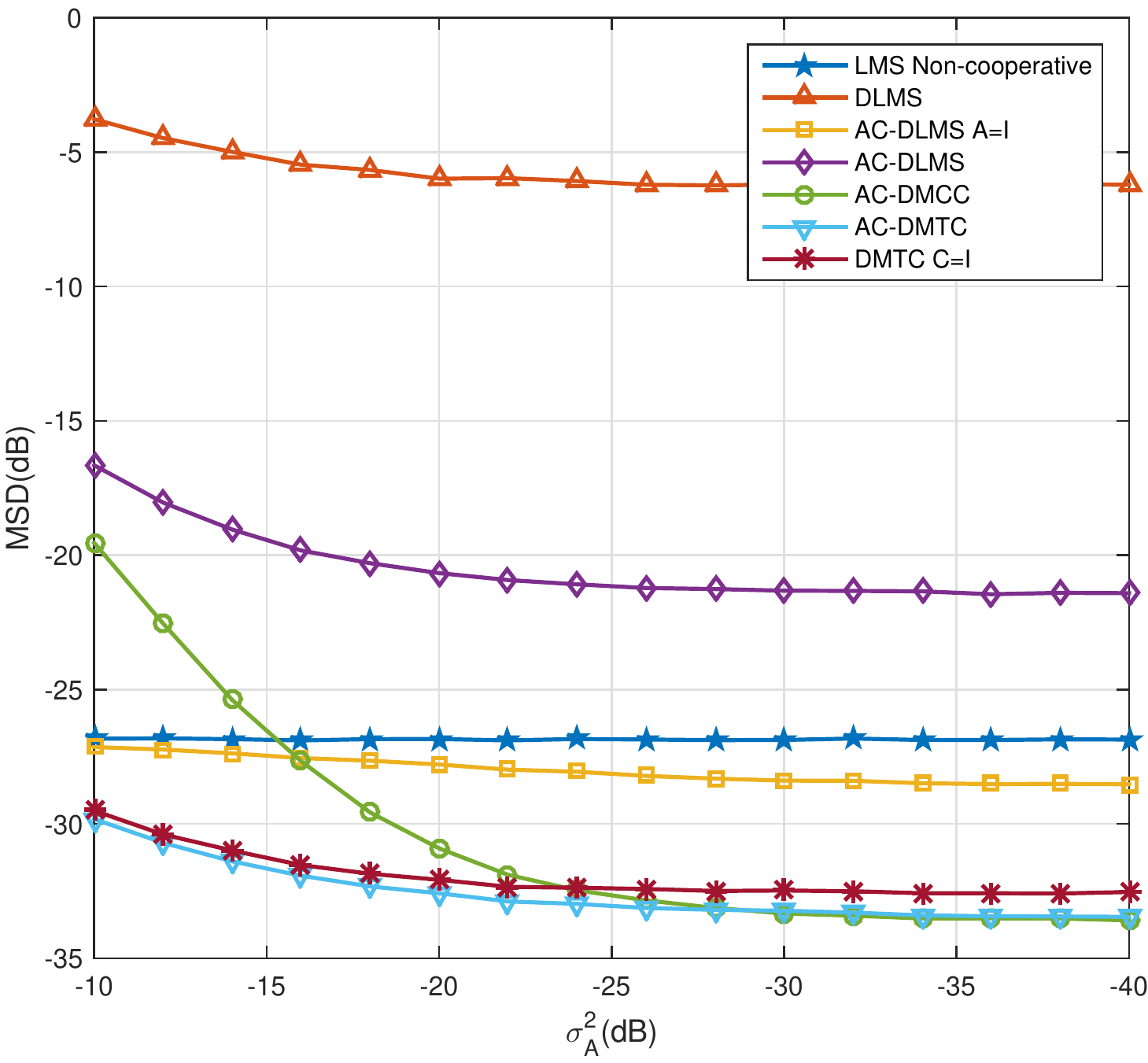}\hspace{0.15in}}
\subfigure[GMM link noise with different $\sigma_B^2$ ($\sigma_A^2=0.01,c=0.01$)]{
\includegraphics[width=0.28\linewidth]{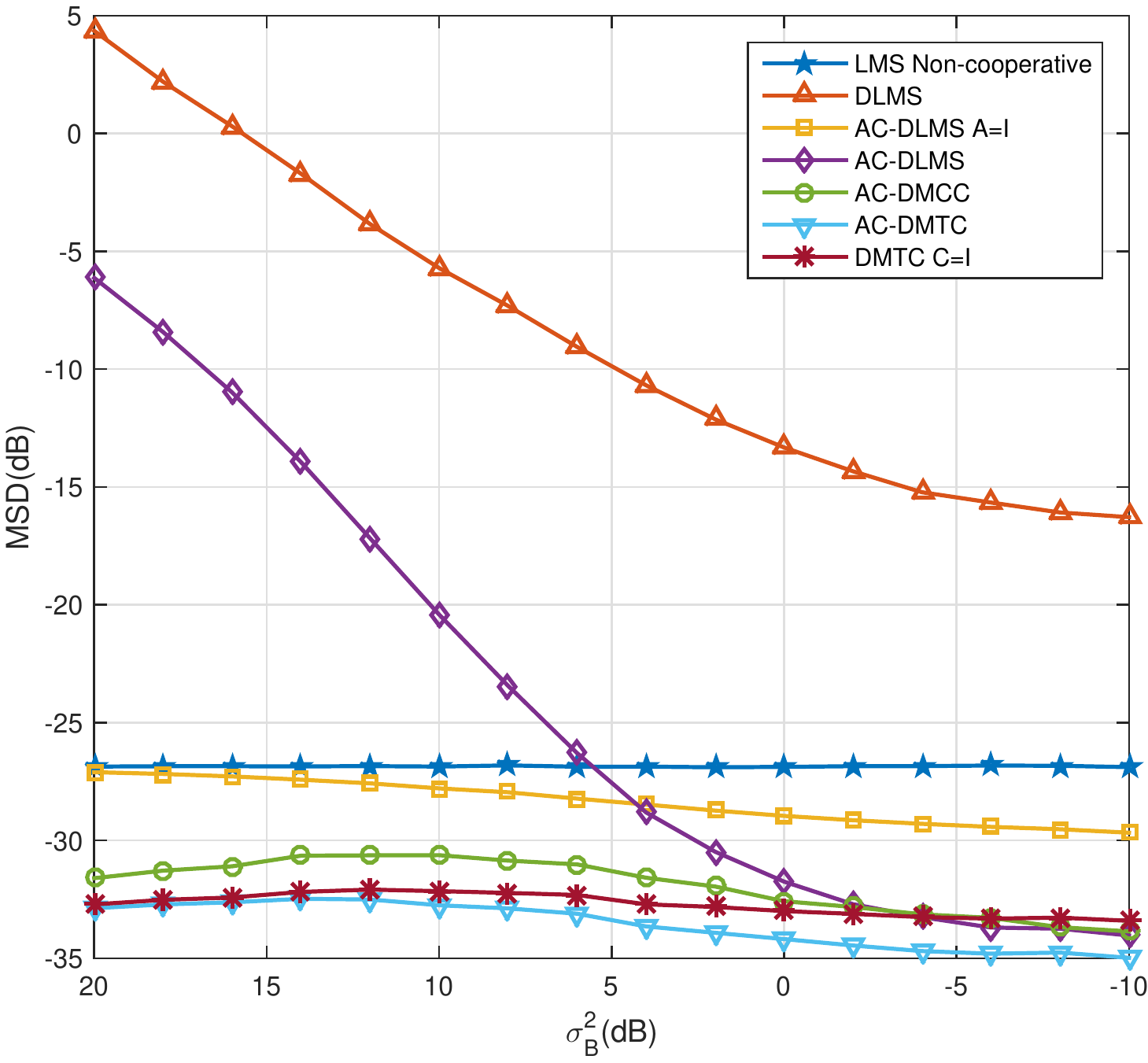}\hspace{0.15in}}
\caption{MSD with different noise distributions}
\label{fig2}
\end{figure*}

Several algorithms are used for comparison including DLMS based diffusion algorithms (DLMS, AC-DLMS, AC-DLMS without data sharing ($\boldsymbol{A}\!\!=\!\!\boldsymbol{I}$)) and DMTC based diffusion algorithms (AC-DMTC, DMTC with only data sharing ($\boldsymbol{C}\!\!=\!\!\boldsymbol{I}$)). 
Non-cooperative LMS and diffusion MCC (DMCC) \cite{Ma2016Diffusion} are also added for comparison. For adaptive combination rule, the forgetting factor $\chi$ is set to 0.05, and $\epsilon$ is set to $10^{-6}$. For DMTC based diffusion algorithms, without mentioned, $\zeta^2_{lk}$ are all set to $10^4$ for $i<100$ and $5\sigma_{A}^2$ for $i>100$. The kernel widths of DMCC are set to $10^4$ for $i<100$ and $5(\sigma_{k}^2+2\sigma_{A}^2)$ for $i>100$ so that it achieves similar measurement of residuals with DMTC at steady state. The MSD is approximated 
over 1000 Monte Carlo runs. For all algorithms, Metropolis weights \cite{lopes2008diffusion} are used in both adaptation and combination step. The initial weight vector is zero vector.
\par First we compare the convergence performance of the algorithms. The observation noise variances $\sigma_{k}^2$ are set to 0.1 for all nodes. All the link noises are set to the same distribution. In the first 2000 iterations, Gaussian link noises with $\sigma_{A}^2=0.04,c=0$ are used. While for $i>2000$, the GMM link noises with $\sigma_{A}^2=0.04,\sigma_{B}^2=10,c=0.01$ are added. The step sizes of each algorithm are adjusted so that all the algorithms have almost the same initial convergence speed. The average learning curves of different algorithms are shown in Fig.\ref{fig1}. One can observe that DMTC based algorithms can achieve better result than others in both Gaussian and non-Gaussian noises. Note that when large outliers occur, the performance of all diffusion methods will degrade.
\par Second, we investigate the performance under different link noises. The algorithm settings are the same as previous simulation. The MSD curves in terms of different noise variances $\sigma_{A}^2$ and $\sigma_{B}^2$ are shown in Fig.\ref{fig2}. One can see that the proposed AC-DMTC algorithm achieves the best performance in all noise environments. In particular, Fig.\ref{fig2}(a) shows that using EIV model can achieve better performance especially when the link noise variance is large. 
Fig.\ref{fig2}(b) confirms the robustness of correntropy, and Fig.\ref{fig2}(c) illustrates that adaptive combination rule can efficiently alleviate the bad influence of large outliers.

\begin{figure}[tb]
\centering
\includegraphics[width=0.85\linewidth]{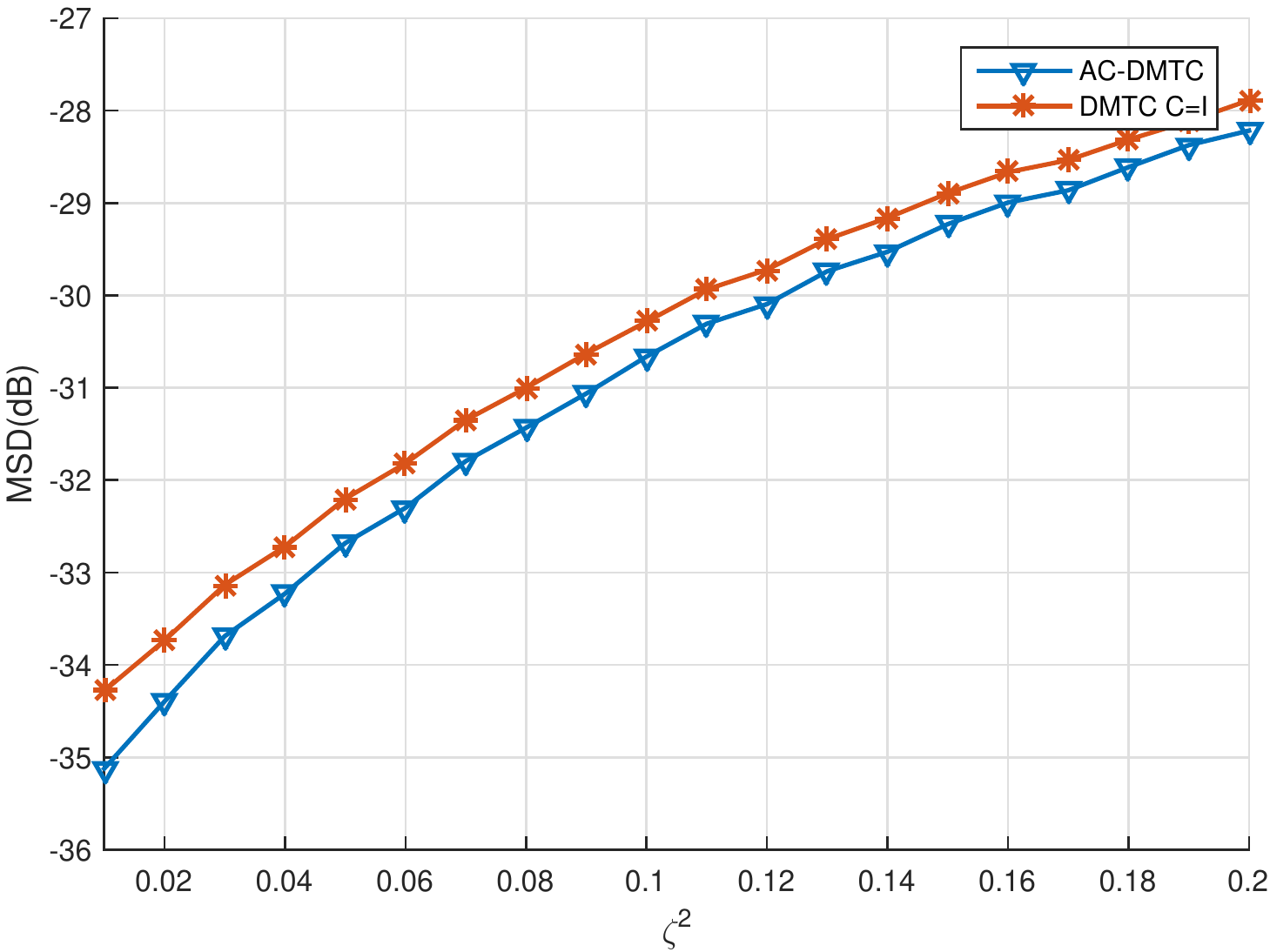}
\caption{MSD with different kernel parameters}
\label{fig3}
\end{figure}

\par Finally, the sensitivity of the kernel parameter $\zeta_{lk}$ is investigated. The GMM noise is set the same with the first simulation. The $\zeta^2_{lk}$ at $i>100$ are all set to the same value $\zeta^2$. The MSD curves versus different $\zeta^2$ are shown in Fig.\ref{fig3}. It can be seen that the MSD increases as the $\zeta$ grows large. Moreover, a small $\zeta$ will slow down the convergence speed, thus one should select a proper value of $\zeta$ to make a trade off between faster convergence speed and lower MSD.

\section{Conclusion}

In this work we propose a new algorithm called diffusion maximum total correntropy (DMTC), which aims to improve both the accuracy and robustness for distributed estimation over network with noisy links. Taking advantages of correntropy and EIV model, the proposed algorithm is theoretically unbiased in Gaussian noise, and can efficiently handle noise in presence of large outliers. Further, the adaptive combination rule is also utilized. Simulation results confirm that the proposed algorithm can achieve excellent performance in both Gaussian and non-Gaussian noises.

%

\bibliographystyle{IEEEtran}
\bibliography{refs}

\end{document}